# FUSeg: The Foot Ulcer Segmentation Challenge

Chuanbo Wang*[1], Amirreza Mahbod[2], Isabella Ellinger[2], Adrian Galdran[3], Sandeep Gopalakrishnan[1], Jeffrey Niezgoda[4], Zeyun Yu[1]

[1] University of Wisconsin-Milwaukee, USA
[2] Institute for Pathophysiology and Allergy Research, Medical University of Vienna, Austria
[3] Department of Computing and Informatics, Bournemouth University, UK
[4] Advancing the Zenith of Healthcare (AZH) Wound and Vascular Center, USA

**Abstract**
Acute and chronic wounds with varying etiologies burden the healthcare systems economically. The advanced wound care market is estimated to reach $22 billion by 2024. Wound care professionals provide proper diagnosis and treatment with heavy reliance on images and image documentation. Segmentation of wound boundaries in images is a key component of the care and diagnosis protocol since it is important to estimate the area of the wound and provide quantitative measurement for the treatment. Unfortunately, this process is very time-consuming and requires a high level of expertise. Recently automatic wound segmentation methods based on deep learning have shown promising performance but require large datasets for training and it is unclear which methods perform better. To address these issues, we propose the Foot Ulcer Segmentation challenge (FUSeg) organized in conjunction with the 2021 International Conference on Medical Image Computing and Computer Assisted Intervention (MICCAI). We built a wound image dataset containing 1,210 foot ulcer images collected over 2 years from 889 patients. It is pixel-wise annotated by wound care experts and split into a training set with 1010 images and a testing set with 200 images for evaluation. Teams around the world developed automated methods to predict wound segmentations on the testing set of which annotations were kept private. The predictions were evaluated and ranked based on the average Dice coefficient. The FUSeg challenge remains an open challenge as a benchmark for wound segmentation after the conference.

*Corresponding author

## 1. Introduction

Chronic nonhealing and acute wounds represent a difficult challenge to healthcare systems, affecting millions of patients globally [1]. In the United States, the cost prediction for wound care treatment is estimated to be between $28.1B and $96.8B [2]. In contrast to acute wounds, chronic wounds fail to predictably proceed through the common phases of healing in an orderly and timely fashion, thus hospitalization and additional care are necessary but increase the cost for health services in billions annually [3]. Outpatient costs ($9.9–$35.8 billion) of wound care are reported to be significantly higher than inpatient costs ($5.0–$24.3 billion) [4]. However, the access and quality of care to chronic wound patients are often limited in primary and rural healthcare settings. A vast majority of chronic wound patients usually have other health conditions such as diabetes, obesity, and circulation problems. The shortage of well-trained wound care clinicians also worsens the situation.

Accurate and fast measurement of the wound area is critical to the management and evaluation of chronic wounds to monitor the wound healing process and to determine future interventions. Clinicians typically measure wounds by their length, width, and depth in clinical practices. Wound length and width are measured with a ruler guide and the depth is measured with Q-tips. However, manual measurement is time-consuming and often inaccurate which can cause a negative impact on patients. Reliable automatic wound segmentation from images would enable automation of the wound area measurement as well as efficient data entry into the electronic medical record to enhance patient care. With the recent advances in deep learning, image-based semantic segmentation of wounds offers an expressive characterization of the wound. However, deep-learning methods impose an even larger burden of manual effort than most manual measurements since they need to be trained on large datasets with pixel-wise labeled images.

A few wound datasets are publicly available. The Diabetic Foot Ulcer (DFU) Challenge dataset [5] contains 15,683 DFU image patches. Unfortunately, the images are not labeled with segmentation masks. The Medetec wound dataset [6] consists of free stock images of all types of open wounds such as venous leg ulcers, arterial leg ulcers, pressure ulcers (pressure sores), diabetic ulcers. Despite covering almost all wound types, the dataset only contains 341 unlabeled images, which is far from enough to train deep learning models. WoundDB [7] contains 188 sets of wound photographs where each set includes four modalities, RGB image, thermal image, stereo image, and depth map. The wounds are fully labeled with outlined boundaries. Like Medetec, the number of images in WoundDB is not sufficient for training deep segmentation models.

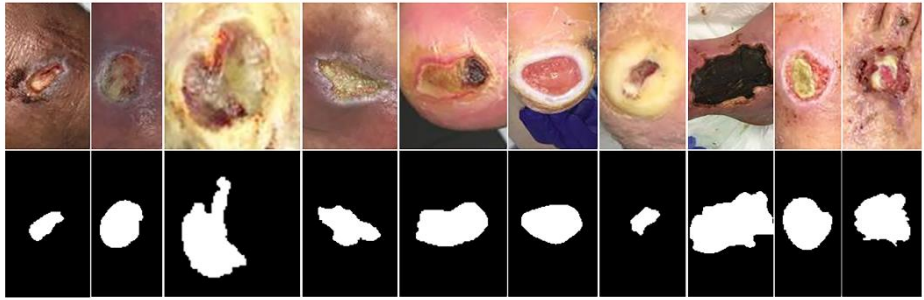

**Figure 1**. The challenge dataset consists of 1,210 foot ulcer images taken from 889 patients. The first row contains the raw images collected. The second row consists of segmentation mask annotations.

## 2. Materials and Methods

### 2.1. Dataset

We collaborated with the Advancing the Zenith of Healthcare (AZH) Wound and Vascular Center to build a chronic wound dataset. It was collected over 2 years from October 2019 to April 2021 at the center and contains 1,210 foot ulcer images taken from 889 patients during multiple clinical visits. The raw images were taken by Canon SX 620 HS digital camera and iPad Pro under uncontrolled illumination conditions, with various backgrounds. The images (shown in Figure 1) are randomly split into 3 subsets: a training set with 810 images, a validation set with 200 images, and a testing set with 200 images. Of course, the annotations of the testing set are kept private. We confirm that the data collected were de-identified and in accordance with relevant guidelines and regulations and the patient's informed consent is waived by the institutional review board of the University of Wisconsin-Milwaukee.

Deep learning models learn the annotations of the training dataset during training. Thus, the quality of annotations is essential. Automatic annotation generated with computer vision algorithms is not ideal when deep learning models are trained to learn how human experts recognize the wound region. In our dataset, initial annotation masks were firstly manually proposed for the raw images. These proposals were further reviewed and modified by wound care specialists with over 20 years of wound care experience and medical assistants from the collaborating wound clinic who followed up the patient visits. For tricky cases and disagreements, the annotation team and our lab would sit together and make final decisions on the annotations. During the annotation process, granulation, slough, and eschar tissues were annotated as wounds.

Currently, only foot ulcer images were annotated and included in the dataset as these wounds tend to be smaller than other types of chronic wounds, which makes it easier and less time-consuming to manually annotate the pixel-wise segmentation masks. In the future, we plan to create larger image libraries to include all types of chronic wounds, such as venous leg ulcers, pressure ulcers, and surgery wounds as well as non-wound reference images. The AZH Wound and Vascular Center, Milwaukee, WI, had consented to make our dataset publicly available.

## 2.2. Challenge Design

### 2.2.1. Infrastructure and Timeline

The homepage of our challenge was hosted on the Grand-Challenge platform and the dataset is stored in our repository on GitHub. We created an individual webpage to present the leaderboard. The challenge was announced on February 24, 2021, and the training set was published on March 9, 2021. In early July we made the testing set available for participants to perform sanity checks and the submission was closed on July 15, 2021. The testing results and rankings for FUSeg2021 were published on August 16, 2021, 2 months before our workshop in MICCAI 2021. After the conference, FUSeg remains an open challenge and the dataset remains publicly available.

### 2.2.2. Submission and Evaluation

Each participating team was asked to submit a docker container that contains their algorithm and the prediction code. We generated segmentation predictions using the code and re-produced the results to be evaluated on our GPU server and all results were evaluated under the same software and hardware environment. The Dice coefficient [8] was used to quantize the performances of submitted algorithms:

$$\text{Dice} = \frac{2 \times True\ positives}{2 \times True\ positives + False\ negtives + False\ positives}$$

where *True positives*, *False negatives*, and *False positives* represent the corresponding number of pixels. Over 100 researchers registered for the FUSeg2021 challenge and 8 teams successfully submitted their algorithms. In section 2.3, we provide an in-depth description of the methods of the top three ranked teams.

Wang

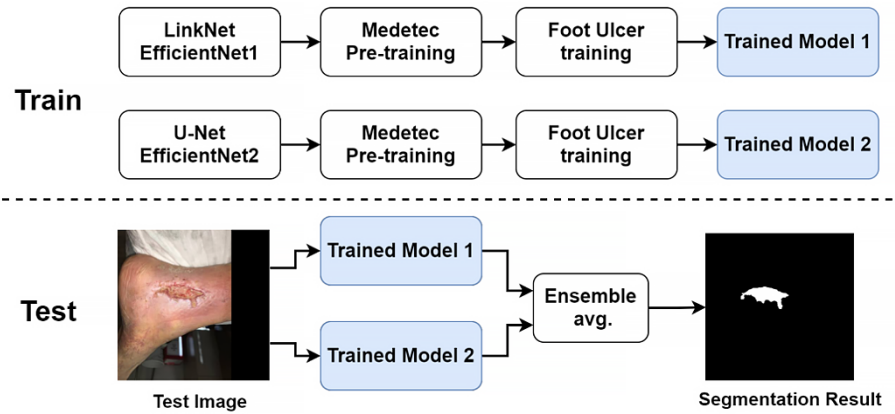

**Figure 2**. The architecture of Mahbod's model. In the training phase, LinkNet and U-Net are used to obtain two trained models whose predictions are ensembled in the testing phase.

### 2.3. Top Three Submissions

#### 2.3.1. First Place

This submission [9] was made by Amirreza Mahbod, Rupert Ecker, and Isabella Ellinger of the Medical University of Vienna.

**Architecture**

Two CNN models were used, namely U-Net [10] and LinkNet [11]. Instead of using the models in their plain forms, pre-trained CNNs were used in the decoder parts of the models. For LinkNet, a pre-trained EfficientNetB1 model [12] was used, and EfficientNetB2 model [12] was utilized for U-Net. As shown in Figure 2, the entire Medetec dataset was also used for pre-training. To train the models, random scaling, random rotations, vertical and horizontal flipping, and brightness and contrast shifts were used as augmentation techniques as suggested in [13]. Each model was trained for 80 epochs with a learning rate (LR) scheduler that reduced the LR by 90% after every 25 epochs. The initial LR was set to 0.001. The batch size was set to 4 and we used the full-size images to train the models. Adam optimizer and a combination of Dice loss and Focal loss are also adopted for model training. For each dataset, fivefold cross-validation was exploited and the best models based on the segmentation scores of the validation sets were saved to be used in the inference phase.

**Ensemble**

To boost the segmentation performance, three distinct ensembling strategies were used, namely 5-fold cross-validation, test time augmentation (TTA), and result fusion from the two exploited models. Instead of using the entire training set to train a single

Wang

model, it was divided randomly into five subsets. Then, five sub-models were trained based on the derived subsets (i.e., for each of the sub-models, four subsets were used for training and the hold-out set was used for validation). In the inference phase, the test images were sent to all five derived sub-models and then took the average over the results. As shown in former studies for various image segmentation or classification tasks [14], TTA can boost the overall performance. Therefore, TTA was used in the inference phase for better segmentation performance. 0, 90, 180, and 270-degree rotations, as well as horizontal flipping, were applied in TTA. Since two distinct models (LinkNet with EfficientNetB1 backbone and UNet with EfficientNetB2 backbone) were trained, their results were fused in the inference phase for a given test image. And the prediction probability masks were fused by averaging from the two models as shown in Figure 2.

**Post-processing**

To form the final segmentation masks for the test images, first, the fused prediction probability vectors were binarized using a 0.5 threshold. Two post-processing steps were also applied, namely filling the holes, and removing very small detected objects, with identical settings as described in [15].

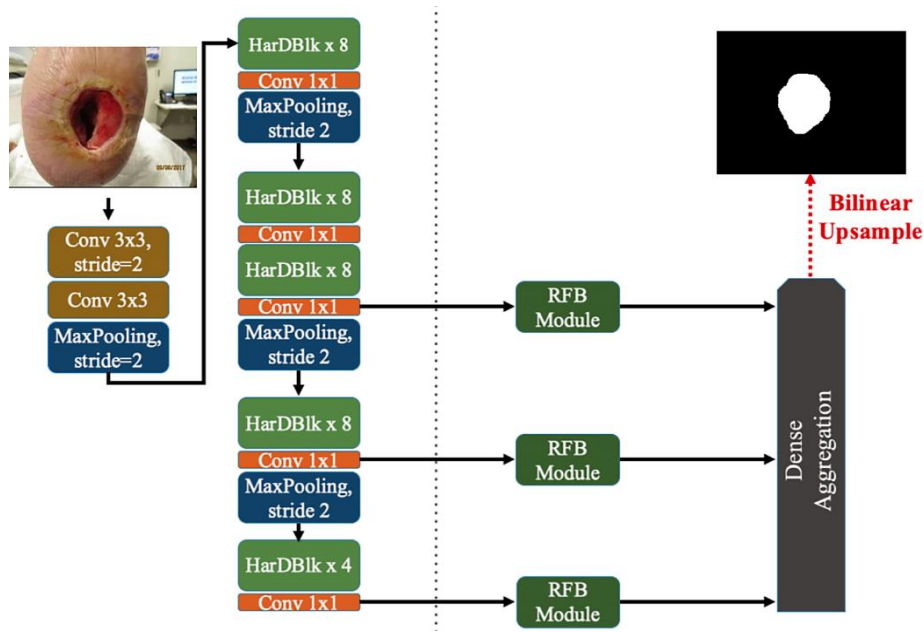

**Figure 3**. HarDNet-MSEG architecture. [16]

### 2.3.2. Second Place

This submission was made by Yichen Zhang from Huazhong University of Science and Technology. In this approach, the Harmonic Densely Connected Network (HarDNet-MSEG) [16] is applied to the wound dataset. As shown in Figure 3, the architecture consists of an encoder backbone and a decoder.

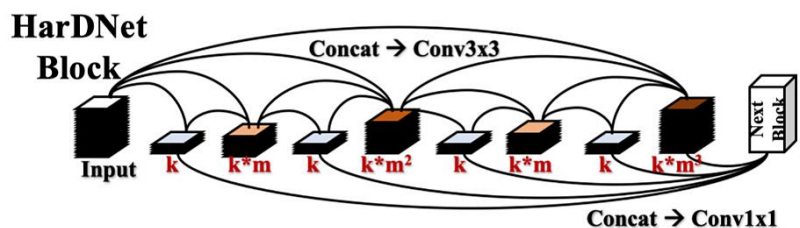

**Figure 4**. Demonstration of a HarDNet block. [16]

Specifically, HarDNet-68 is adopted as the encoder. It consists of repeated application of HarDNet blocks (Figure 4), batch normalizations (BN), and max pooling layers. The arrangement of the layers follows the standard Conv-BN-ReLU order to enable the folding of BN. For the layer distribution, instead of concentrating on stride-16 that is adopted by most of the CNN models, stride-8 is used to have the most layers in the HarDNet-68 that improves the local feature learning benefiting small-scale object awareness. In each HardNet block, layer k is connected to layer $k–2^n$ if $2^n$ divides k, where n is a non-negative integer and $k–2^n \geq 0$. Under this connection scheme, once layer $2^n$ is processed, layer 1 through $2^n–1$ can be flushed from the memory reducing the concatenation cost significantly.

The decoder is implemented as a cascaded partial decoder [17]. It found out that the shallow features with high resolution occupy computing resources, and the deep information can also represent the spatial details of the shallow information relatively well. Thus, the shallow features are discarded and allow for more computing on the deeper layers' features. At the same time, the aggregation of feature maps at different scales can be achieved by adding appropriate convolution and skip connections.

During training, multi-scale training and standard image augmentations are applied. The Dice coefficient is evaluated to be 87.57% on the testing set of the wound dataset ranking our proposed method the 2nd place on the leaderboard.

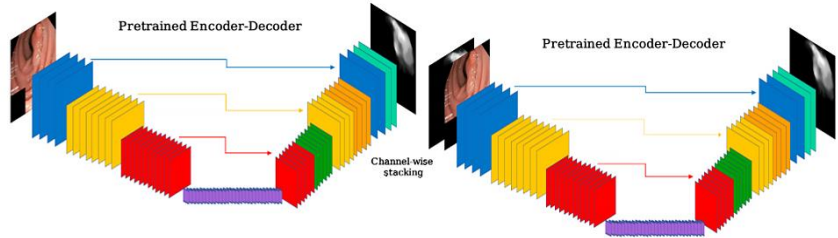

**Figure 5**. Pre-trained Double Encoder-Decoder Network [18]. The second network receives input as the original raw image concatenated with the prediction of the first network. This allows the second network to better focus on the region of interest in the image, i

### 2.3.3.   Third Place

This submission [18] was made by Adrian Galdran of the University of Bournemouth. In this work, a framework for semantic segmentation was proposed based on the sequential use of two encoder-decoder networks The architecture consists of two segmentation networks stacking in a sequential manner, where the second network receives as input the concatenation of the prediction from the first one with the original frame, as shown in Figure 5. This way, the output of the first network acts as an attention mechanism that provides the second network with a map of interesting locations on which the second network should focus. Double encoder-decoders are a direct extension of encoder-decoder architectures in which two encoder-decoder networks are sequentially combined. Denoting by x the input RGB image, $E^{(1)}$ the first network, and $E^{(2)}$ the second network. In a double encoder-decoder, the output $E^{(1)}(x)$ of the first network is provided to the second network together with x so that it can act as an attention map that allows $E^{(2)}$ to focus on the most interesting areas of the image:

$$E(x) = E^{(2)}\left(x, E^{(1)}(x)\right), \tag{1}$$

where x and $E^{(1)}(x)$ are stacked so that the input to $E^{(2)}$ has four channels instead of the three channels corresponding to the RGB components of x. There are some choices to be made in this framework, specifically about the structure of the encoder and decoder sub-networks within $E^{(1)}$ and $E^{(2)}$. Note that $E^{(1)}$ and $E^{(2)}$ do not need to share the same architecture, although in this work we selected U-Net for both sub-networks.

Wang

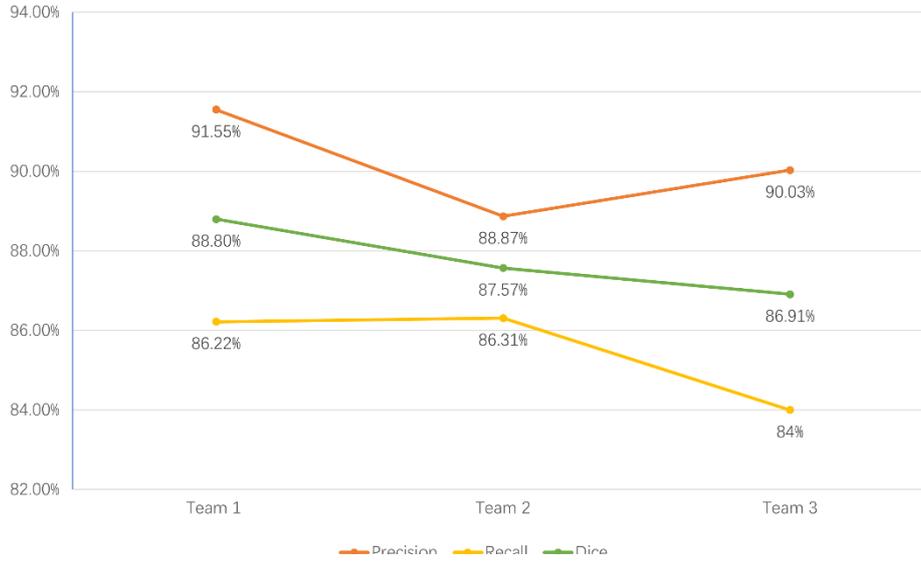

**Figure 6**. Performance evaluation of the top three submitted algorithms.

## 3. Results

In this section, we report the performances of the top three submitted algorithms. Besides the Dice coefficient, we also calculate the precision and recall to demonstrate the performances in detail. As shown in Figure 6, the ensemble network (Team 1) consists of a U-Net and a LinkNet achieves the highest precision of 91.55% and the highest dice coefficient of 88.8%. HarDNet-MSEG (Team 2) was evaluated to have the highest recall of 86.31% but the lowest precision of 88.87%, hurting its overall dice score to be ranked as second place. The lowest recall and dice score was evaluated from the double encoder-decoders model submitted by Team 3.

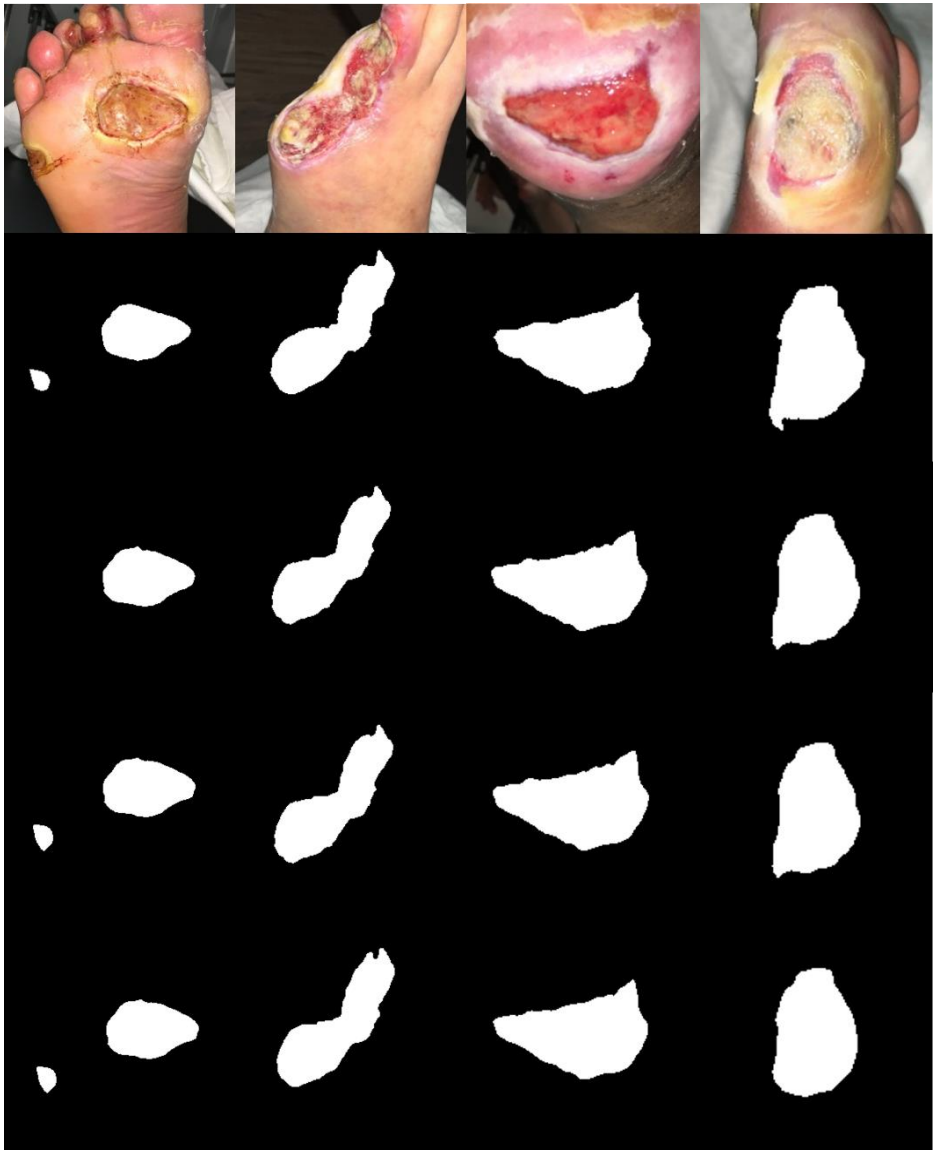

**Figure 7**. Demonstration of the predicted segmentation masks from the top three submissions. Each row in the top-down order are the original images, ground truth masks, Mahbod, Zhang, and Galdran, respectively.

Wang

## 4.  Discussion

**Overall Segmentation Performance**

From the annotation experience, we learned the wound segmentation problem was difficult for even human experts who followed up the entire healing process. One of the challenges is that boundaries between epithelial tissue and granulation tissue are often ambiguous due to their similar color and the fact that epithelial tissue forms on top of granulation tissue.

The overall performance of the submitted algorithms is promising. In general, all submitted algorithms were capable of segmenting the wound region quite well (Figure 7), with Dice scores of over 80%. The best Dice score of 88.8% and a surprising precision of 91.55% were reached by the top team. Segmenting the relatively bigger wounds with clear boundaries worked well when the wound beds are properly cleansed and the dead tissue are removed, and reasonably well for cases where infection, slough, or other impediments were present. Segmenting small isolated areas of the wound with ambiguous boundaries was the most difficult task.

**The U-Net architecture**

The performance of vanilla U-Net was evaluated to have a Dice score of 80% on our dataset. However, U-Net is still a popular option for the participants as a subnetwork in ensemble networks or the encoder part of their architecture. For example, U-Net is used as one of the subnetworks pre-trained with carefully selected models in the submission by Mahbod. Galdran also makes use of U-Net by stacking two sequentially where the input of the second network is the concatenation of the original input and the prediction from the first network. Among the submissions, we also find other variants of U-Net such as U-Net with atrous spatial pyramid pooling layers [19] and U-Net with the addition of residual connections [20].

**Ensemble networks**

When comparing the two groups of models (ensemble networks vs single networks), we observed that fusing predictions from different networks always outperforms the single networks in our dataset. This observation agrees with the theory of ensemble learning, ensembles that combine multiple networks tend to yield more generalized and robust results when there is significant diversity among the networks. In this challenge, we see two of the main ensemble learning methods, stacking and boosting. Both algorithms perform better than the individual networks, namely U-Net and LinkNet when applied to solve the same problem.

Wang

## 5.  Conclusions

In this paper, we presented the Foot Ulcer Segmentation challenge, which remains an "open" challenge to serve as a challenging benchmark in the semantic segmentation of wounds after MICCAI 2021. We built a large pixel-wise annotated wound image dataset that is manually labeled by experts and evaluated the submitted wound segmentation methods. Although wound segmentation is a difficult problem, our results suggest:

- Current state-of-the-art algorithms can accurately segment the wound area, with Dice scores reaching 88.8% and precision reaching 91.55%.
- From the predictions generated by the submitted algorithms, we observed the challenges in distinguishing between epithelial tissue and granulation tissue and segmenting small isolated wound regions.
- We also observed the superiority of ensemble networks over individual networks applied to our dataset.

## Acknowledgments

I wish to acknowledge the great help provided by the doctors and medical assistants in the AZH Wound and Vascular Center. The assistance provided by my colleague Behrouz Rostami in the proposal of initial annotations was greatly appreciated. I would also like to show my deep appreciation to my supervisors for their patient guidance, encouragement, and advice.

## About the authors

**Chuanbo Wang** is a graduate student in the Department of Computer Science, University of Wisconsin-Milwaukee, Milwaukee, Wisconsin, United States. **Amirreza Mahbod** is a research scientist of the Institute for Pathophysiology and Allergy Research, Medical University of Vienna, Vienna, Austria. **Isabella Ellinger** is an Associate Professor of the Institute for Pathophysiology and Allergy Research, Medical University of Vienna, Vienna, Austria. **Adrian Galdran** is a researcher of the Department of Computing and Informatics, Bournemouth University, UK. **Sandeep Gopalakrishnan**, Ph.D., DAPWCA, is an Assistant Professor in College of Nursing, University of Wisconsin-Milwaukee, Milwaukee, WI. **Jeffrey Niezgoda**, MD, FACHM, MAPWCA, CHWS, is the President, Chief Executive Officer, and Corporate Medical Director of AZH Wound and Vascular Center, Milwaukee, WI. **Zeyun Yu**, Ph.D., is a Professor in Computer Science and Biomedical Engineering at the University of Wisconsin-Milwaukee, Milwaukee, WI.

Wang

Wang